\documentclass[runningheads]{llncs}
\usepackage{graphicx}
\usepackage{float}
\usepackage{setspace}
\usepackage{multirow}
\usepackage{enumitem}
\usepackage{cite}
\usepackage{xcolor}
\usepackage{array}
\usepackage[caption=false]{subfig}
\newcolumntype{C}[1]{>{\centering\arraybackslash}p{#1}}
\newcolumntype{M}[1]{>{\centering\arraybackslash}m{#1}}
\makeatletter
\renewcommand\section{\@startsection{section}{1}{\z@}%
                       {-8\p@ \@plus -4\p@ \@minus -4\p@}%
                       {6\p@ \@plus 4\p@ \@minus 4\p@}%
                       {\normalfont\large\bfseries\boldmath
                        \rightskip=\z@ \@plus 8em\pretolerance=10000 }}
\renewcommand\subsection{\@startsection{subsection}{2}{\z@}%
                       {-8\p@ \@plus -4\p@ \@minus -4\p@}%
                       {6\p@ \@plus 4\p@ \@minus 4\p@}%
                       {\normalfont\normalsize\bfseries\boldmath
                        \rightskip=\z@ \@plus 8em\pretolerance=10000 }}
\renewcommand\subsubsection{\@startsection{subsubsection}{3}{\z@}%
                       {-4\p@ \@plus -4\p@ \@minus -4\p@}%
                       {-1.5em \@plus -0.22em \@minus -0.1em}%
                       {\normalfont\normalsize\bfseries\boldmath}}
\makeatother
\begin{document}

\title{CyberSecurity Challenges for Software Developer Awareness Training in Industrial Environments}
\titlerunning{CyberSecurity Challenges for Industrial Software Developers}

\author{
  Tiago Gasiba\inst{1,2} \and
  Ulrike Lechner\inst{2}  \and
  Maria Pinto-Albuquerque\inst{3}
}
\authorrunning{T. Gasiba et al.}
\institute{
  Siemens AG, Munich, Germany\\
  \email{tiago.gasiba@siemens.com}
  \and
  Universität der Bundeswehr München, Munich, Germany\\
  \email{tiago.gasiba@unibw.de} \email{ulrike.lechner@unibw.de}
  \and
  Instituto Universitário de Lisboa (ISCTE-IUL), ISTAR-IUL, Lisboa, Portugal\\
  \email{maria.albuquerque@iscte-iul.pt}
}

%\author{
%  xxxxx xxxxxx\inst{1,2} \and
%  xxxxxx xxxxxxx\inst{2}  \and
%  xxxxx xxxxxxxxxxxxxxxxx\inst{3}
%}
%\authorrunning{x. xxxxxx et al.}
%\institute{
%  xxxxxxx xxx xxxxxxx xxxxxxx\\
%  \email{xxxxxxxxxxxx@xxxxxxxxxxx}
%  \and
%  xxxxxxxxxxx xxx xxxxxxxxxx xxxxxxxx xxxxxxx xxxxxxx\\
%  \email{xxxxxxxxxxxx@xxxxxxxx} \email{xxxxxxxxxxxxxx@xxxxxxxx}
%  \and
%  xxxxxxxxx xxxxxxxxxxxxx xx xxxxxx xxxxxxxxxxxx xxxxxxxxxx xxxxxxx xxxxxxxx\\
%  \email{xxxxxxxxxxxxxxxxx@xxxxxxxxxxxx}
%}

\maketitle

\begin{abstract}
Awareness of cybersecurity topics facilitates software developers to produce secure code. This awareness is especially important in industrial environments for the products and services in critical infrastructures. In this work, we address how to raise awareness of software developers on the topic of secure coding. We propose the "CyberSecurity Challenges", a serious game designed to be used in an industrial environment and address software developers' needs. Our work distills the experience gained in conducting these CyberSecurity Challenges in an industrial setting. The main contributions are the design of the CyberSecurity Challenges events, the analysis of the perceived benefits, and practical advice for practitioners who wish to design or refine these games.


\end{abstract}
\section{Introduction}
\label{sec:introduction}
%\vspace{-.32cm}

Over the last years, the number of industrial security-related incidents, e.g., reported by the ICS-CERT~\cite{ICS_CERT}, has been steadily increasing.
When malicious parties exploit security vulnerabilities present in products and services, the outcome of its exploitation has serious negative consequences for society, the customers, and the company that produced the software. Think, e.g., of critical infrastructures as the grid, transportation, or production lines: a security vulnerability in the code may cause interruptions in service quality or cause safety issues for society or individual customers when critical machinery fails. Several efforts can be made to increase the level of security in critical infrastructures.
These efforts include, among others: analysis of threat and risks, implementing a secure software development lifecycle process, deployment of static application security testing tools, code reviews, and training.

This paper addresses the software vulnerabilities through awareness training of software developers in the industry, based on a serious game: the CyberSecurity Challenges (CSC).
Serious Games are games that {\it are designed for a primary purpose other than pure entertainment} \cite{2016_Doerner_Serious_Games}. The serious game "CyberSecurity Challenges" (CSC) aims at raising awareness of secure coding topics among industrial software engineers.
In this game, software developers are trained to spot security vulnerabilities in software and write secure code. i.e., code that is free from known vulnerabilities and adheres to secure coding policies.
Previous work introduced the CyberSecurity Challenges from a theoretical point-of-view \cite{gasiba_re19,Gasiba2020_QUATIC} and focused on particular aspects \cite{Gasiba2020_CyberICPS}.
The current work extends previous publications by a presentation of a unified view on the design process, tailoring to the industry's needs and the perceived usefulness of the CSC events. Our results are based on data from several CSC events held in the industry from 2017 to 2020. 
As such, the main contributions of this work are:
\begin{itemize}
    \item {\bf CSC Artifact}: consolidated view of the design and deployment of CSCs, based on results from thirteen events held in an industrial context, and
    \item {\bf CSC Evaluation}: analysis of results from industry events covering the following aspects: adequacy of CSC as a means to raise secure coding awareness, impact of CSC on software developers, and success factors for CSC events.
\end{itemize}

This paper aims to guide practitioners who wish to develop or refine a software developer awareness training in an industrial context, provide a solid reference to the research community who wishes to address serious games for the industry, and close the existing literature gap.
This work is organized as follows.
In the following section, we will give a summary of what CSC games are and describe its origin.
Section \ref{sec:related_work} presents previous work related.
In Section \ref{sec:Method}, the research method and research questions are introduced.
The unified view of the CSC artifact is presented in Section \ref{sec:csc_industry}.
Section \ref{sec:results} presents a summary of the survey results, together with critical discussions.
Finally, Section \ref{sec:conclusions} concludes the paper with an outlook of the work and next steps.

\section{Cybersecurity Challeges at a glance}
The CyberSecurity Challenges (CSCs) are a serious game, designed to raise awareness for cybersecurity topics among industrial software engineers. 
A CSC game consists of several challenges designed to raise awareness on secure coding guidelines and secure coding on software developers.
These challenges are oriented towards improving the defensive skills of the participants.
Defensive challenges are challenges that help the players write code that has no (known) vulnerabilities and adheres to secure coding guidelines.

The Capture-the-Flag genre was the original inspiration for the game. Capture-the-Flag (CTFs) are associated with offensive skills, e.g., system penetration, and reverse engineering, and they can often last hours or even days~\cite{mirkovic2014class}.
Unlike CTF games, which teach the participants to attack and break into systems, CSC focus on improving skills to write and develop secure code.
These games thus have no intention to cause any harm or inspire unlawful actions.
The challenges are composed of C, C++, Java, and Web exercises.
The focus on these programming languages and genre inspiration is rooted in internal demand for training and internal decisions taken in the company where CSC is developed.
Thus, the games are designed to match software developers' interests and organizations' needs for developer training.
This interest can be motivated by several factors, e.g., the need to show due diligence and certification purposes.

The CSC event is delivered individually (Standalone) or after a workshop on secure coding (Workshop). In both cases, the duration of the event is designed to fit a single working day.
During the game, the participants solve secure coding challenges related to secure coding guidelines, either individually or as part of a team.
Although the challenges can include an offensive part (e.g., on how malicious parties exploit systems), the main focus and emphasis of the challenges is on developing secure software, i.e., on the defensive perspective.
For each solved challenge, points are awarded, and the winner of the game is the one with the highest amount of points.
Participants to the event can have either a background in a single programming language or be mixed, e.g., both C and Web developers.

\section{Related Work}
\label{sec:related_work}
Although several methods exist to deal with software vulnerabilities, e.g., requirements engineering and code reviews, we focus on awareness training for software developers.
Several previous studies indicate that software developers lack secure programming awareness and skills~\cite{Assal2019,gitlab_2019,tahaei2019survey}.
In 2020, Bruce Schneier, a well-known security researcher, and evangelist stated that {\it less than 50\% of software developers can spot security vulnerabilities in software} \cite{Schneier2020}. His comment adds to a discussion on secure coding skills: In 2011, Xie et al.~\cite{Xie2011} did several interviews with 15 senior professional software developers in the industry with an average of 12 years of experience. Their study has shown a disconnect between software security concepts and their role in their jobs.
Awareness training on Information security is addressed in McIlwraith \cite{McIlwraith2006}, which looks at employee behavior and provides a systematic methodology and a baseline on implementing awareness training.
In their work, Stewart et al. \cite{Stewart2012} argue that communicators, e.g., trainers, must understand the audiences' constraints and supporting beliefs to provide an effective awareness program.

There is a stream of literature on compliance with security policies, which deals with employees in general and not with software developers specifically. This stream of literature explores many reasons why people do not comply with IT-security policies. The unified framework by Moody et al.~\cite{moody2018toward} summarizes the academic discussion on compliance with IT-security policies. Empirical findings include that neither deterrence nor punishment such as e.g., public blame, works to increase compliance. However, increasing IT-security awareness increases the level of compliance \cite{Stewart2012}. In their seminal review article, Hänsch et al.~\cite{2014_Benenson_Defining_Security_Awareness} define IT-security awareness in the three dimensions: {\it Perception} (knowledge of existing software vulnerabilities), {\it Protection} (knowing the existing mechanisms - best practices - that avoid software vulnerabilities), and  {\it Behavior} (knowledge and intention to write secure code). The concept of IT-security awareness is typically used in IT security management contexts, and we use this concept to evaluate our work. While these findings are for the compliance of employees with IT-security policies and awareness of IT security, little empirical research is done on IT-security awareness in software development and what makes software developers comply with security policies in software development.

Graziotin et al.~\cite{Graziotin2018} show that \textit{happy developers are better coders}, i.e., produce higher quality code and software.
Their work suggests that by keeping developers happy, we can expect that the code they write has a better quality and, by implication, be more secure.
Davis et al.\cite{Davis2014} show, in their construct, that cybersecurity games have the potential to increase the overall happiness of software developers. Their conclusions support our approach to use a serious game approach to train software developers in secure coding.
Awareness games are a well-established instrument in information security and are discussed in de-facto standards as the BSI Grundschutz-Katalog \cite{2016_Grundschutz_Katalog} (M 3.47, Planspiele) as one means to raise awareness and increase the level of security. Frey et al.~\cite{Maria2019} show both the potential impact of playing cybersecurity games on the participants and show the importance of playing games as a means of cybersecurity awareness. They conclude that cybersecurity games can be a useful means to build a common understanding of security issues. Rieb et al. \cite{Rieb2018} provide a review of serious games in cybersecurity and conclude that there are many approaches. However, only a few have an evaluation of their usefulness and are available beyond the immediate context of a consulting or cyber-security company. The games listed mainly address information security rather than secure coding. Documented and evaluated games are \cite{2016_Beckers_Serious_Game} and \cite{Rieb2018}.

Capture-the-flag is one particular genre of serious games in the domain of Cybersecurity \cite{Davis2014}. Game participants win flags when they manage to solve a task. Forensics, cryptography, and penetration testings are skills necessary for solving tasks and capturing flags. They are considered fun, but there are hardly any empirical results on these games' effects on participants' skill levels.
The present work uses serious games to achieve the goal of {\it raising secure coding awareness of software developers in the industry}. 
Previous work on selected design aspects and a smaller empirical basis on the CSC includes \cite{Gasiba2020_RankingSCG,Gasiba2020_PlayerProfile,gasiba_re19,Gasiba2020_CyberICPS,Gasiba2020_CyberICPS_Journal,Gasiba2020_QUATIC,Gasiba2020_TrustCOMM}.

\section{Method}
\label{sec:Method}
%The serious game entitled "CyberSecurity Challenges" is serious game that has been developed to address the inherent need to raise awareness of industrial software developers on the topic of secure coding.
The design science paradigm, according to Hevner \cite{2004_hevner_design_science}, Baskerville and Heje \cite{baskerville2010explanatory} guides our research in the industry.
Design and evaluation of designs in iterative approaches are an integral part of design research: this article presents our design after 13 CSC events and the evaluation of the design. The events took place from 2017 to 2020, with more than 200 game participants.

%\begin{spacing}{.7}
  \begin{table*}[htb]
    \renewcommand{\arraystretch}{1.05}
    \scriptsize
    %\footnotesize
    %\small
    \centering
    \caption{Overview of Cybersecurity Challenges Events}
    \label{tbl:csc_events}
    \vspace{-1em}
    \begin{tabular}{|C{.7cm}|C{1.7cm}|C{1.8cm}|C{1.6cm}|C{1.6cm}|C{.7cm}|C{2.3cm}|}
        \hline
        {\bf No.} & {\bf Date} & {\bf Type} & {\bf Focus} & {\bf Where} & {\bf NP} & {\bf Data collection} \\
        \hline
        \hline
        1 & Nov. 2017 & Standalone   & Mixed  & Germany & 11 & SSI\\
        \hline
        2 & May. 2018 & Standalone   & Web   & Germany & 12 & SSI \\
        \hline
        3 & Jul. 2018 & Standalone   & Web   & Germany & 6 & SSI\\
        \hline
        4 & Jul. 2018 & Standalone   & Mixed  & Germany & 30 & SSI\\
        \hline
        5 & Sep. 2018 & Standalone   & Web   & Germany & 16 & SSI\\
        \hline
        6 & Aug. 2019 & Workshop & C/C++   & China & 14 & Survey\\
        \hline
        7 & Aug. 2019 & Workshop & Web   & China & 15 & Survey\\
        \hline
        8 & Sep. 2019 & Workshop & Web   & Germany & 7 & Survey\\
        \hline
        9 & Oct. 2019 & Workshop & C/C++   & Turkey  & 23 & Survey\\
        \hline
        10 & Jun. 2020 & Standalone   & C/C++   & Online & 15 & Survey$^*$\\
        \hline
        11 & Jul. 2020 & Standalone   & C/C++   & Online & 21 & Survey$^*$\\
        \hline
        12 & Jul. 2020 & Standalone   & C/C++   & Online & 20 & Survey$^*$\\
        \hline
        13 & Jul. 2020 & Standalone   & C/C++   & Online & 15 & Survey$^*$\\
        \hline
    \end{tabular}
    {\\\hspace{\textwidth}
             {\scriptsize {\bf NP}: No. of players, {\bf SSI}: semi-structured interview, {\bf (*)} for survey description see \cite{Gasiba2020_CyberICPS}}
             }
  \end{table*}
%\end{spacing}

Table~\ref{tbl:csc_events} summarizes the CSC events. CSC games were designed in three design cycles: 1) Initial Design (events 1-5), 2) Refinement (events 6-9) and 3) Sifu/Online (events 10-13).
The CSC events participants were all software developers specializing in web technologies and the C/C++ programming language. The events took place mostly in Germany but also in China and Turkey.
The players' age ranged from 25 to 60, the background industry of the participants was critical infrastructures, in particular, industry automation (50.85\%), energy (37.29\%), and healthcare (11.86\%), the overall number of years of work experience was as follows: one year (13.7\%), two years (11.0\%), three years (19.2\%), four years (6.8\%) and five or more years (49.3\%).
Regarding the average number of security training over the previous five years, the results are as follows: Germany - 3.57, China - 2.10, and Turkey 1.50.

According to the first and second design cycles, the evaluation of these CSC events is structured according to the following research questions. For analysis of the survey results concerning the Sifu platform, we refer the reader to~\cite{Gasiba2020_CyberICPS}.
\begin{itemize}[leftmargin=+.43in]
    \item[ {\bf RQ1:}] To what extent are CSC adequate to raise awareness about secure coding?
    \item[ {\bf RQ2:}] What is the impact that CSC workshops have on the participants?
    \item[ {\bf RQ3:}] Which factors are considered essential for a successful CSC event?
\end{itemize}

To address these research questions, the authors have conducted semi\- structured interviews (SSI) \cite{newcomer2015conducting}, and developed a small survey.
The semi-structured interview questions were asked to the participants, one after another in a round-the-table. The participants' answers were recorded on paper.
The semi-structured interviews were performed during the first design cycle and were part of the feedback round after the CSC event.
They were based on the following questions: a) "what went well and you would you like to keep" and b) "what did not go well and would you like to change".
These questions gave a good insight and allowed us to improve later versions of the game. They were also fundamental for requirements elicitation (see~\cite{gasiba_re19}).

%\begin{spacing}{.7}
  \begin{table*}[http]
    \renewcommand{\arraystretch}{1.05}
    \scriptsize
    %\footnotesize
    %\small
    \centering
    \caption{Survey 1: Questions}
    \label{tbl:survey:survey_1}
    \vspace{-1em}
    \begin{tabular}{|p{0.7cm}|p{0.6cm}|p{0.75cm}|p{9.6cm}|}
        \hline
        %% CT -> Construct
        {\bf RQ} & {\bf CT} & {\bf QID} & {\bf~~~~~~~~~~~~~~~~~~~~~~~~~~~~~~~~~~~~~Question} \\
        \hline
        \hline
        %%% RQ1
         \multirow{11}{0.7cm}{~RQ1} & \multirow{3}{0.6cm}{~PE} &  & By participating in this awareness training \underline{\hspace{1.2cm}} \\
         \cline{3-4}
         &&Q1.1  &  ~~~~I learned new techniques and principles of secure software development\\
         \cline{3-4}
         && Q2.1 &  ~~~~I understand the possible consequences of a security breach \\
         
         \cline{2-4}
         & \multirow{5}{0.6cm}{~BE}&Q3.1 &  ~~~~I feel that I am prepared to handle secure coding related issues at work \\
         \cline{3-4}
         &&Q4.1 &  ~~~~I understand the need to have secure development life-cycle activities \\
         \cline{3-4}
         &&Q5.1 &  ~~~~I feel more prepared to work with static code analysis tools (e.g. SAST) \\
         \cline{3-4}
         &&Q6.1 &  ~~~~I know how to use the information about secure coding guidelines \\
         \cline{3-4}
         &&Q7.1 & Focusing on the challenges improves my practical secure coding skills \\
         
         \cline{2-4}
         &\multirow{3}{0.6cm}{~PR}&Q8.1 &  I have learned about new issues that I would like to check in my own code\\
         \cline{3-4}
         &&Q9.1 &  I know where I can find more information about secure coding guidelines \\
         \cline{3-4}
         &&Q10.1 &  I understand the importance of secure coding guidelines \\         

         %%%%%% RQ2
         \hline
         \multirow{3}{0.7cm}{~RQ2} & \multirow{3}{0.6cm}{~~--} &Q11.1 & The learning goals of the challenges were clearly explained \\
         \cline{3-4}
         &&Q12.1 & CSC games help me to understand the need to develop secure software \\
         \cline{3-4}
         &&Q13.1 &  The help from the coaches was adequate \\

        %%%%%%% RQ3
        \hline
        \multirow{8}{0.7cm}{~RQ3}&\multirow{8}{0.6cm}{~~--}&Q14.1 & I want to learn about new tools, even if I do not use them at work\\
        \cline{3-4}
        &&Q15.1 & I prefer to solve challenges sequentially rather, even if it takes too much time \\       
        \cline{3-4}
        && Q16.1& Working in teams is better than working individually on the challenges \\
        \cline{3-4}
        &&Q17.1 & I like the fact that different kinds of challenges are presented \\
        \cline{3-4}
        &&Q18.1 & I prefer challenges that address the same problem from different point-of-views \\
        \cline{3-4}
        &&Q19.1 & I prefer challenges that are related with my work environment \\
        \cline{3-4}
        &&Q20.1 & I prefer challenges that are based on real-life examples \\
        \cline{3-4}
        &&Q21.1 & I prefer challenges that can be systematically solved with some tool \\
        \hline
    \end{tabular}
    {\\\hspace{\textwidth}
                     {\scriptsize {\bf RQ}: Research Question, {\bf CT}: Construct , {\bf QID}: Question Identifier, {\bf PE}: Perception, ~~~~~~~~~~~~~~~~~~~~~~~~~~~~~~~~~~~~~~~~~~~~~~~~~~~~~ {\bf BE}: Behavior, {\bf PR}: Protection}~~~~~~~~~~~~~~~~~~~~~~~~~~~~~~~~~~~~~~~~~~~~~~~~~~~~~~~~~~~~~~~~~~~~~
                     }
  \end{table*}
%\end{spacing}

The survey was administered to the CSC participants, in the refinement cycle, after completion of the event. The survey consisted of an online survey. Participation in the SSI and the survey was opt-in. Furthermore,  all participants consented to participate in research, and the collected data was anonymized.
We have used a more formal survey methodology to evaluate the game's usefulness concerning the level of awareness and the skills in secure coding.
Table~\ref{tbl:survey:survey_1} shows the questions that were asked in the survey and the related research questions. The survey used a five-point Likert scale of agreement with the following mapping: strongly disagree (1), disagree(2), neutral (3), agree (4), and strongly agree (5).
RQ1 addresses the aspect of the usefulness of the CSC artifact, and the corresponding survey questions are based on the three dimensions of awareness, as defined by Hänsch et al. \cite{2014_Benenson_Defining_Security_Awareness}: Perception (PE), Protection (PR) and Behavior (BE) (cf. Sec. Related Work).
The questions for RQ2 focus on clarity of the description of the challenges, the coaches' role during the game, and the general motivation of training secure coding. These questions address the design of CSC games and events.
RQ3 questions address the challenges and their relation to software developers' everyday work practices in the industry. The survey questions for RQ2 and RQ3 are based on the authors' experience in industrial software engineering, feedback from CSC evaluations of events 1 to 5, and various discussions with colleagues.
%A mapping between the survey questions and the theoretical constructs are also shown in Table~\ref{tbl:survey:survey_1}.
All the collected data were processed using the statistics package RStudio 1.2.5019.
Availability of the gathered data is provided in the same authors' included references and on a forthcoming publication.
% This forthcoming publication is the PhD thesis

\section{Design of the CyberSecurity Challenges}
\label{sec:csc_industry}
In this section, we present the design of the CyberSecurity Challenges for industrial software developers. The sub-sections provide a detailed overview of the architecture, the schedule, and the design of challenges. The results presented in this section distill the experience obtained through the three design cycles of the CSC games, i.e., of the thirteen CSC events.

\subsection{Architecture}
%A CyberSecurity Challenge event is a one-day event in which 10 to 30 software developers from the industry participate. Teams of players are presented with several challenges. Upon solving a challenge, a flag is awarded. This flag can be submitted to a dashboard to redeem for points. At the end of the event, the team with the most points wins the CSC game. For each event, the kind and number of challenges and other event elements are tailored to the participants and their secure coding topics. This section presents the general architecture of the game infrastructure and the pragmatic topics of organizing CSC events.

\begin{figure}[http]
    \centering
    \includegraphics[width=.75\columnwidth]{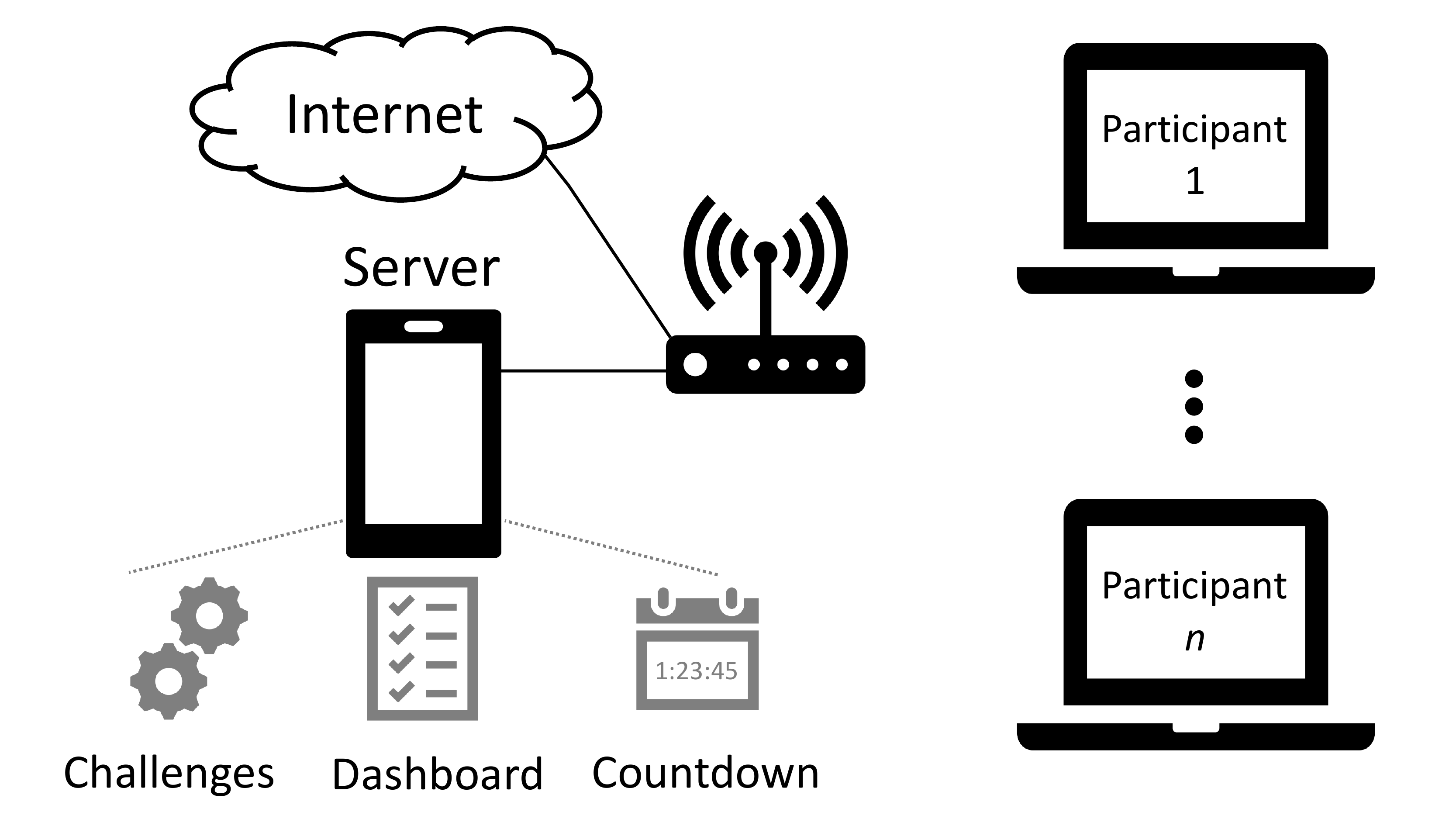}
    \caption{Architecture of CyberSecurity Challenges infrastructure}
    \label{fig:CSC_Architecture}
\end{figure}

Figure~\ref{fig:CSC_Architecture} shows the architecture of CSC infrastructure. Each participant accesses the challenges through a computer. A server hosts the applications that run the game logic, a "countdown" clock, and a dashboard that records individual players and teams' progress. The dashboard uses the open-source CTFd \cite{chung} project. A description of the challenges will be given in the following.

%\subsubsection{CSC Game and Event}
%The CSC game itself can be embedded in a workshop context. Note that we refer to the game part as a CSC game and the game embedded in the whole context as CSC event. In the first ten instances, we collected experiences with Standalone CSC games and workshops. The {\it Standalone} game can be deployed as an (internal) online service. The use cases for the standalone version are (1) a refresher on previously acquired knowledge, (2) as a self-evaluation tool for the individual players, and (3) as a recruiting tool by human resources.
%CyberSecurity Challenges can also be offered together with a secure coding training in a workshop. When a workshop is done after a secure coding training, the CSC's goal is to practice and exercise the learned material.
%Based on direct feedback from participants, our experience has shown that software developers highly appreciate playing CSC after a secure coding training. Furthermore, the participants have claimed that the challenges have helped solidify, understand, and practice in real scenarios the concepts discussed during training.

\subsection{CSC Time Schedule}
Table~\ref{tbl:agenda} shows a typical time-plan for the one-day CSC event consisting of seven blocks: 1) welcome, 2) team building, 3) introduction, 4) main event, 5) winner announcement, 6) feedback and 7) walk-through.

%The different agenda blocks have emerged from the experience gathered from the CSC events and the players' feedback.
The last block, the walk-through, was not initially planned and is the direct result of players feedback --- the participants preferred to dedicate one hour of the main event to provide final explanations and closure on selected exercises.
The authors decided to place the feedback and survey before the walk-through to increase the chance of collecting feedback from the participants.

\begin{spacing}{.9}
  \begin{table*}[http]
    \renewcommand{\arraystretch}{0.99}
    \scriptsize
    %\footnotesize
    %\small
    \centering
    \caption{Agenda for a one-day cybersecurity challenges game event}
    \label{tbl:agenda}
    \begin{tabular}{|C{1.25cm}|p{1.8cm}|p{8.85cm}|}
        \hline
        {\bf Duration} & \multirow{1}{*}{\bf ~~~~What} & \multirow{1}{*}{\bf ~~~~~~~~~~~~~~~~~~~~~~~~~~~~~~~~Description} \\
        %~~~(hrs) & & \\
        \hline
        \hline
        10~min        & {\it Welcome}  & Welcome to participants and accessing CSC infrastructure   \\
        \hline
        20~min  & {\it Team building} & Participants select partners and build teams that will play against each other\\
        \hline
        30~min  & {\it Introduction}  & Challenge types are presented. One challenge in each category is solved in order to show the participants how the game works\\
        \hline
        320~min & {\it Main event} & Game is open and teams are free to play the game. They are responsible for defining their own strategy for time-out (e.g. for lunch break).\\
        \hline
        10~min      & {\it Winner}        & Game is closed and teams can no longer submit points to the dashboard. Winning team is announced. A brief review of the game-play is done together with the participants.\\
        \hline
        30~min       & {\it Feedback}  & Participants are asked to fill out a survey about the game. Additionally, discussions with players is held in short non-systematic interviews. Main points of discussions is recorded for later analysis.\\
        \hline
        60~min      & {\it Walk-through}  & Participants are shown solution to the exercises they considered most difficult. These exercises are solved together in interaction with all the participants. Discussion on how to solve the challenge is highly encouraged. \\
        \hline
    \end{tabular}
  \end{table*}
\end{spacing}

%Note that in the industrial setting, giving feedback or filling out a survey cannot be made mandatory.
The duration of similar training events ranges from several days \cite{mirkovic2014class} (less common) to a single day \cite{SANS642} (more common). Note that the first CTF is done in academia, while a commercial provider does the latter. Additionally, a difference to typical Capture-the-Flag events are the two agenda items {\it Introduction} and {\it Walk-through}.

\subsection{Defensive Challenges}
The primary focus of the CSC game's challenges are Web and C/C++.
In contrast to C/C++, for the web challenges, it was decided not to focus on a single programming language or framework since many of these programming languages and frameworks are in everyday use in the company where the CSC game was developed. In this case, we chose a generic approach based on the Open Web Application Security Project - OWASP~\cite{OWASP}.
The challenges' design took two approaches: 1) based on open-source components and 2) design of own challenges. The first approach was used in the Refinement design cycle, while the second approach in the Sifu/Online design cycle.
A common approach to the design of the challenges is given in~\cite{Gasiba2020_QUATIC}.
Each challenge is presented to the participants according to the following phases: {\it Phase 1} - introduction, {\it Phase 2} - challenge, and {\it Phase 3} - conclusion.
The types of challenges are: Single-Choice Questions (SCQ), Multiple-Choice Questions (MCQ), Text-Entry Questions (TEQ), Associate-Left-Right (ALR), Code-Snippet Challenge (CsC), and Code-Entry Challenge (CEC). Second, 
Phase 1 presents an introduction to the challenge and sets up the scenario; the main part of the challenge is phase 2; phase 3 concludes the challenge by adding additional text related to secure coding guidelines or additional questions related to phase 2.

\subsubsection{Challenges using Open-Source Components}
\begin{spacing}{.9}
  \begin{table*}[http]
    \renewcommand{\arraystretch}{0.99}
    \scriptsize
    %\footnotesize
    %\small
    \centering
    \caption{Open-Source Tools used for Cybersecurity Challenges}
    \label{tbl:tools}
    \vspace{-1em}
    \resizebox{\textwidth}{!}{
    \begin{tabular}{|p{1.4cm}|p{1.8cm}|p{1.1cm}|p{7.5cm}|}
        \hline
        {\bf ~~~Type} & {\bf ~~~~Project} & {\bf ~Effort} & {\bf ~~~~~~~~~~~~~~~~~~~~~~~~~~~Description} \\
        \hline
        \hline
        Web/Java         & Juice Shop               & Minimal & Insecure web application for training purposes from the OWASP project.  \\
        \hline
        Web/Java  & Java & Medium & Secure coding guidelines dedicated to Java from Carnegie Mellon University   \\[-8pt]
        &SEI-CERT&&\\
        \hline
        Web         & Vulnerable & Medium & REST API containing several vulnerabilities \\[-1pt]
        & API &&\\
        \hline
        C/C++       & MBE                      & Small  & Vulnerable code from RPISEC course at Rensselaer Polytechnic Institute\\
        \hline
        C/C++       & C/C++           & Medium &Secure coding guidelines dedicated to C/C++ from Carnegie Mellon University    \\[-8pt]
        &SEI-CERT&&\\
        \hline
        C/C++       & Vulnerable & High & Vulnerable C/C++ code from NIST (Juliet Set) \\[-1pt]
        & code snippets & & \\
        \hline
    \end{tabular}
    }
  \end{table*}
\end{spacing}
Challenges on secure coding for software developers can be implemented by using and adapting existing open source components.
Since most of the available projects focus on the offensive perspective, the following adaptations are suggested: 1) include an incomplete description on how to solve the challenge, and 2) provide follow-up questions related to secure coding guidelines.
Fig.~\ref{fig:web_challenge_phase_1}-\ref{fig:web_challenge_phase_3} shows an example of a challenge for Web developers using OWASP JuiceShop.
This challenge's learning goal is to understand what SQL injections are and how to identify an SQL injection quickly.
Phase 1 sets the stage for the challenge (Fig.~\ref{fig:web_challenge_phase_1}).
In Phase 2, the player is assisted with how to find the vulnerability, through the textual description, as in Fig~\ref{fig:web_challenge_phase_2}, or also directed by the game coaches.
The last phase consists of an additional question related to the exercise, as shown in Fig~\ref{fig:web_challenge_phase_3}, which directs the player to secure coding guidelines.
Table~\ref{tbl:tools} shows the open-source projects and components in which have been used to design CSC challenges for Web and for C/C++, along with the expected effort required to modify them.
Note that the design of these challenges is based on open source components that include an offensive perspective. Therefore, after the components' adaptation, it is more natural and more accurate to describe these types of challenges as being {\it defensive/offensive}.

%Note that a preliminary presentation of formats and structure has been provided in previous work \cite{Gasiba2020_QUATIC}. Note furthermore that previous work~\cite{gasiba_re19,Gasiba2020_QUATIC} has elaborated that the CSC challenges should focus on defensive rather than offensive security topics. 

\subsubsection{Defensive Challenges using Sifu Platform}
\begin{figure}[http]
    \centering
    %\vspace*{-1em}
    \begin{minipage}{.49\textwidth}
        \centering
        \includegraphics[width=.99\linewidth]{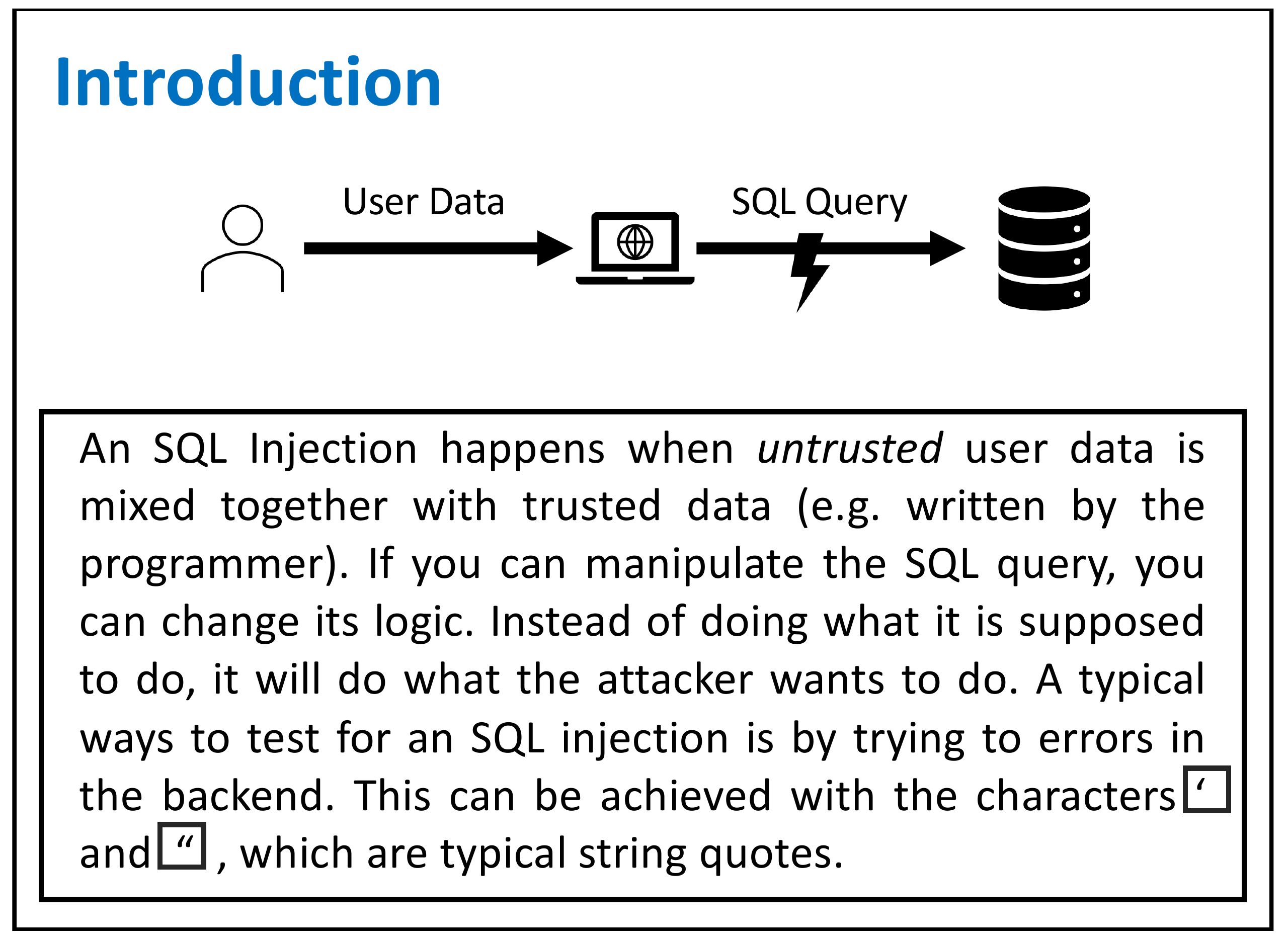}
        \vspace*{-2em}
        \caption{Web Challenge: Phase 1}
        \label{fig:web_challenge_phase_1}
        \vspace*{1em}
    \end{minipage}
    \begin{minipage}{.49\textwidth}
        \centering
        \includegraphics[width=.99\linewidth]{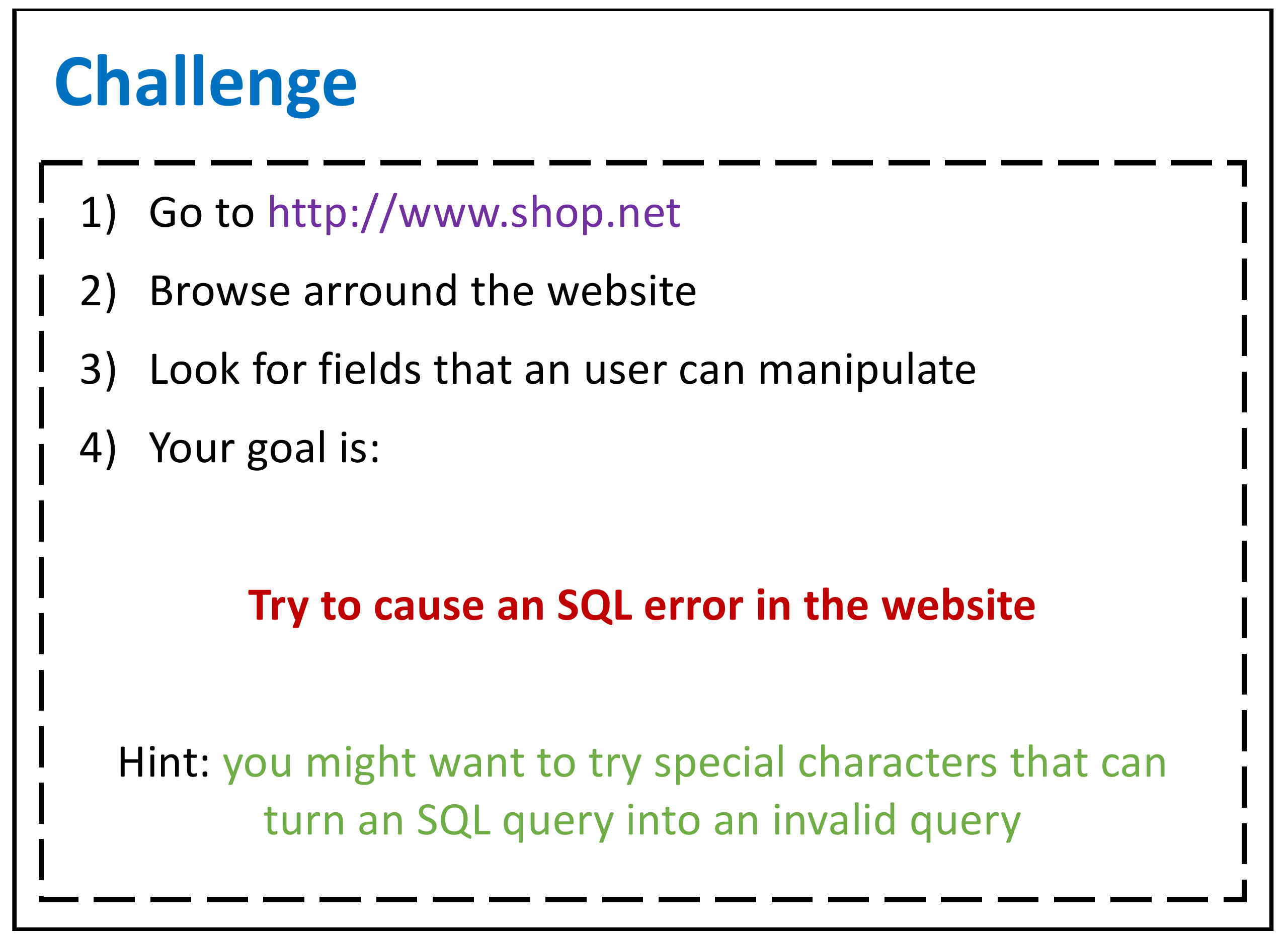}
        \vspace*{-2em}
        \caption{Web Challenge: Phase 2}
        \label{fig:web_challenge_phase_2}
        \vspace*{1em}
    \end{minipage}

    \begin{minipage}{.48\textwidth}
        \centering
        \includegraphics[width=.99\linewidth]{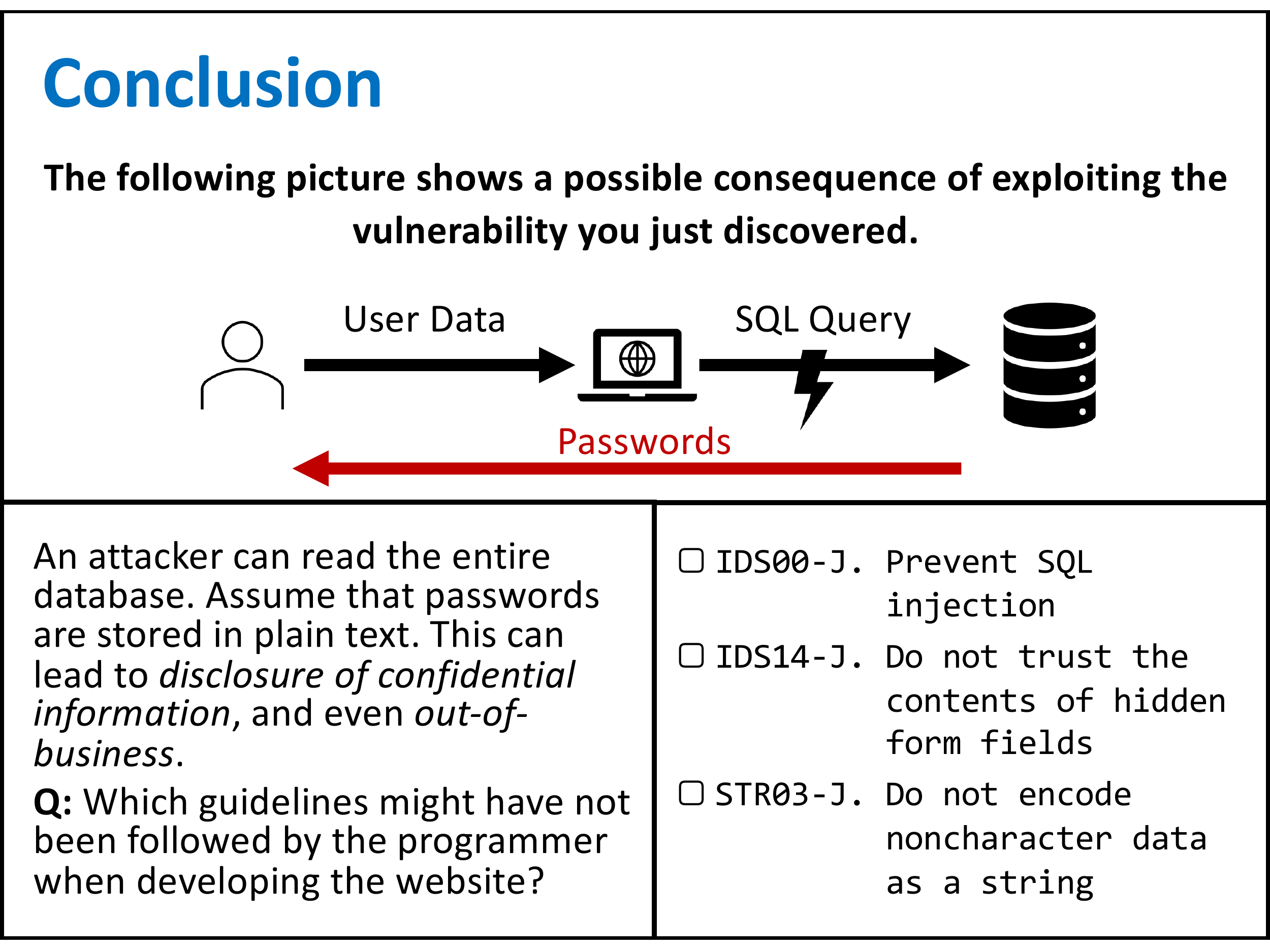}
        \vspace*{-2em}
        \caption{Web Challenge: Phase 3}
        \label{fig:web_challenge_phase_3}
    \end{minipage}
    \begin{minipage}{.48\textwidth}
        \centering
        \includegraphics[width=.99\linewidth]{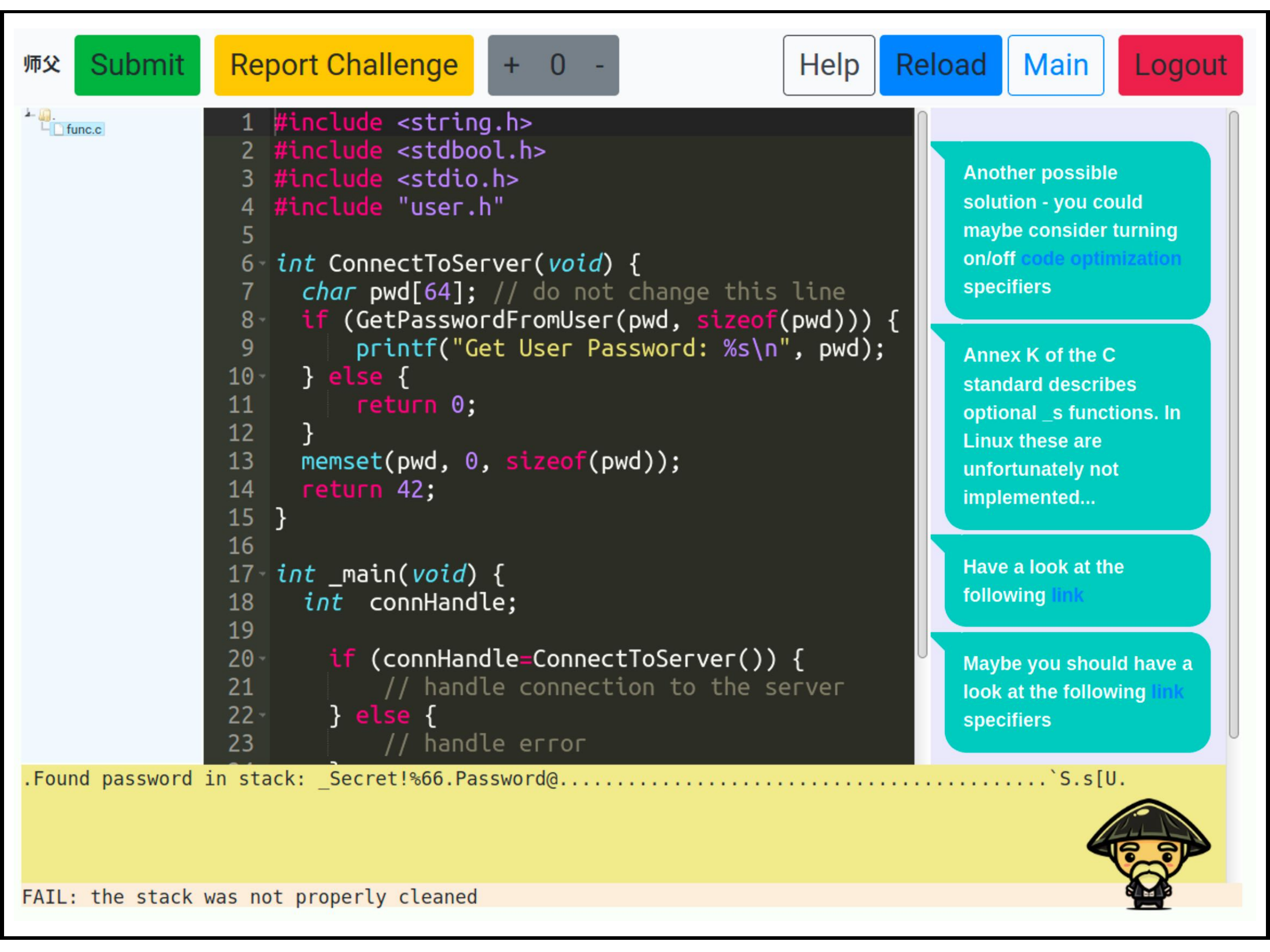}
        \vspace*{-2em}
        \caption{Sifu Platform}
        \label{fig:sifu_platform}
    \end{minipage}
\end{figure}

The Sifu platform hosts code projects containing vulnerabilities in a web application.
The reason to choose a web interface is to avoid that the players need to install any software on their machine, which might be difficult in an industrial setting.
The players' task is to fix the project's source code to bring it to an acceptable solution (therefore focusing on the defensive perspective).
An acceptable solution is a solution where the source code is compliant to secure coding guidelines and does not have known vulnerabilities.
The Sifu platform contains two main components: 1) challenge assessment and 2) an automatic coach.
The challenge assessment component analyses the proposed solution submitted by a player and determines if it is acceptable.
Analysis is based on several tools, e.g., compiler output, static code analysis, and dynamic code analysis.
The automatic coach component is implemented through an artificial intelligence technique that provides hints to the participant when the solution is not acceptable, with the intent to guide the participant to an acceptable solution.
Figure~\ref{fig:sifu_platform} shows the Sifu platform.
Note that only phase 2 is shown in the figure.
The player can browse the different files of the project.
All the hints issued by the automatic coach are available on the right-hand side.
If the player experiences errors when using the platform, these can be reported for later analysis and improvement.
The Sifu platform's main advantage is that the participants do not need to install any software in their machine - a browser with internet or intranet access is sufficient.
However, since untrusted and potentially malicious code will be executed in the platform during the analysis stage, several security mechanisms need to be implemented to guarantee that the players cannot hack it.
These challenges were developed in the Sifu/Online design cycle, and further and detailed information on the implementation is available in~\cite{Gasiba2020_CyberICPS}. For more information about the Sifu platform we also refer the reader to \cite{Gasiba2020_CyberICPS_Journal}.

\section{Results}
\label{sec:results}
This section presents a quantitative analysis of the CSC artifact based on the semi-structured interviews and online survey collected during the design cycles Initial Design and Refinement.

\subsection{Initial Design Cycle ––– CSC 1 to 5}
As discussed in section \ref{sec:Method}, in this design cycle, the participants were asked to provide feedback on what should be kept and what should be changed in the CSC event. The participants were encouraged to discuss what they felt was important openly. These discussions were used to inform the design of future CSC events.
In this cycle, requirements were collected on traits that serious games for software developers in the industry should have.
A summary of the findings is as follows: 1) {\it challenges should focus on the defensive perspective}, 2) {\it challenges should reflect real-world examples}, 3) {\it challenges should be aligned with the work environment}, 4) {\it careful planning in terms of duration should be performed}, and 5) {\it participants should be able to solve challenges without knowledge of extra tools}.
A more in-depth analysis of the feedback and resulting requirements is available in~\cite{gasiba_re19}.

\subsection{Artifact Refinement Cycle ––– CSC 6 to 9}

\begin{figure}[http]
    \centering
    \includegraphics[width=1.0\columnwidth]{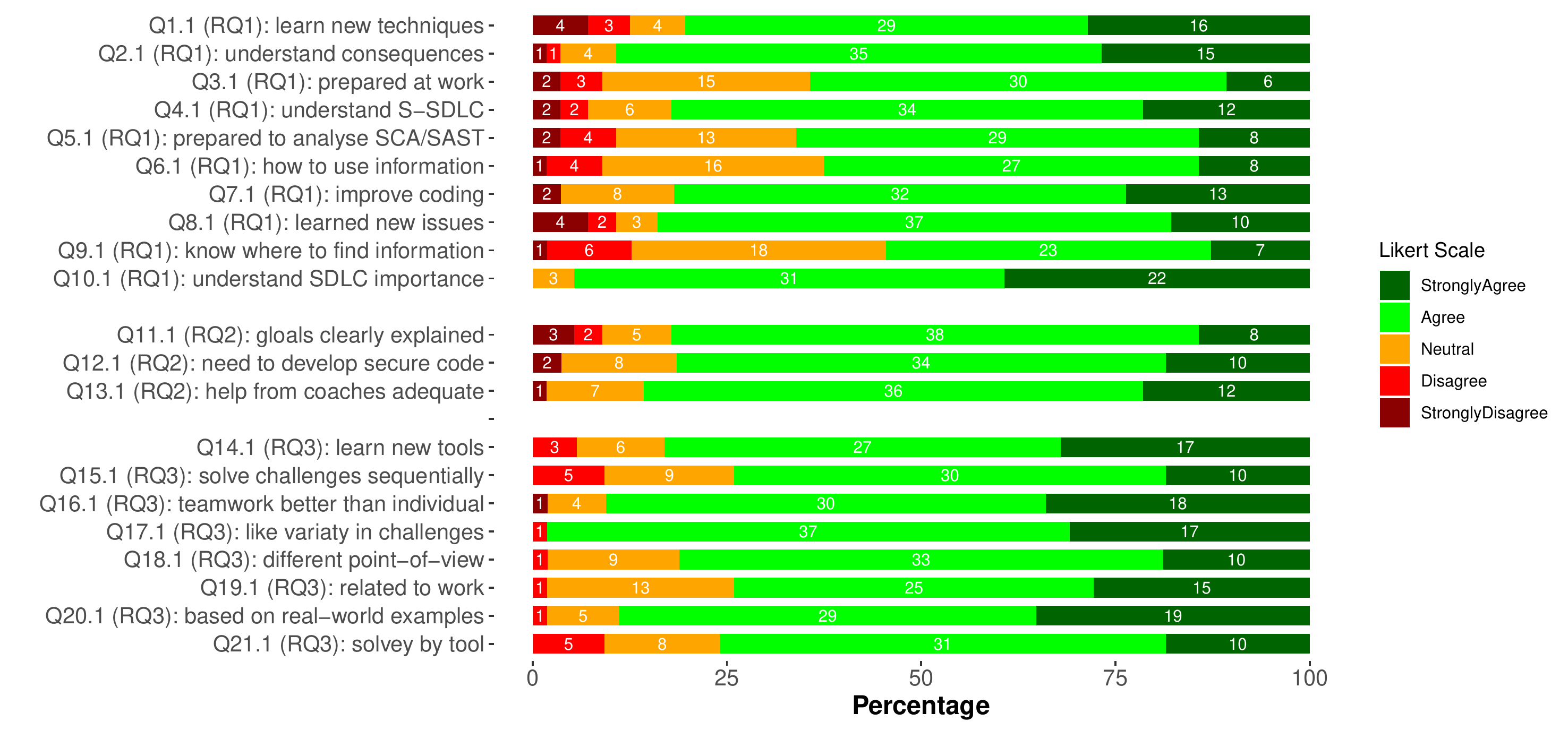}
    \vspace{-2em}
    \caption{Evaluation of Usefulness of CSC Events 6-9}
    \label{fig:survey_1_results}
\end{figure}

Figure~\ref{fig:survey_1_results} shows the overall results of the answers to the survey. The research questions are used to group the results.
We observe an overall agreement on all the survey questions. In particular, considering negative answers (-), neutral answers (N) and positive answers (+), this table shows the following overall results for each research question: $RQ1^-=7.89\%$, $RQ1^N=16.13\%$, $RQ1^+=75.99\%$, $RQ2^-=4.82\%$, $RQ2^N=12.05\%$, $RQ2^+=83.13\%$, $RQ3^-=4.19\%$, $RQ3^N=12.56\%$, and $RQ3^+=83.26\%$.
These results give a good indication that CSC games are suitable as a means to train software developers in secure coding guidelines, as the factors on awareness (RQ1) and impact on participants (RQ2) have high levels of agreement (i.e., higher than 75\%.
However, we observe the difficulty in making every participant happy, in particular, due to the residual values on negative and neutral answers.
Further analysis is required to understand this. Based on our experience, we believe that this fact might be correlated with the participants' previous experience.

\begin{spacing}{.9}
  \begin{table*}[http]
    \renewcommand{\arraystretch}{0.99}
    \scriptsize
    %\footnotesize
    %\small
    \centering
    \caption{Analysis of Research Questions for Survey on CSC Events 6-9}
    \label{tbl:survey_1:rq:analysis}
    \vspace{-1em}
    \begin{tabular}{|p{0.8cm}|p{1cm}|p{0.8cm}|p{0.8cm}|p{0.8cm}|p{0.8cm}|p{0.8cm}|p{0.8cm}|p{0.8cm}|p{0.8cm}|p{0.8cm}|p{0.8cm}|}
        \hline
         & Rank & 1 & 2 & 3 & 4 & 5 & 6 & 7 & 8 & 9 & 10 \\
        \hline
        \hline
        \multirow{2}{0.6cm}{~RQ1}  & W.Avg. & 4.34 & 4.11 & 3.98 & 3.93 & 3.89 & 3.84 & 3.66 & 3.66 & 3.63 & 3.53 \\
        \cline{2-12}
        & QID & Q10.1 & Q2.1 & Q7.1 & Q4.1 & Q1.1 & Q8.1 & Q5.1 & Q6.1 & Q3.1 & Q9.1 \\
        \hline
        \hline
        \multirow{2}{0.6cm}{~RQ2}  & W.Avg. & 4.04 & 3.93 & 3.82 & - & - & - & - & - & - & - \\
        \cline{2-12}
        & QID & Q13.1 & Q12.1 & Q11.1 & - & - & - & - & - & - & - \\
        \hline
        \hline
        \multirow{2}{0.6cm}{~RQ3}  & W.Avg. & 4.27 & 4.22 & 4.21 & 4.09 & 4.00 & 4.00 & 3.85 & 3.83 & - & - \\
        \cline{2-12}
        & QID & Q17.1 & Q20.1 & Q16.1 & Q14.1 & Q18.1 & Q19.1 & Q21.1 & Q15.1 & - & - \\
        \hline
    \end{tabular}
  \end{table*}
\end{spacing}

%\vspace{-1em}

Table~\ref{tbl:survey_1:rq:analysis} shows a ranking of the different survey questions, grouped by research question.
The ranking is performed by sorting the questions based on the average agreement value. In terms of adequacy (RQ1), and impact on the participants (RQ2), the two highest-ranking answers are: to understand the importance of SDLC (Q10), and understand consequences of a breach (Q2) for RQ1, and help from coaches (Q13), and understand the need to develop secure software (Q12) respectively.
The lowest-ranked factors for RQ1 are "find more information" (Q9) and "prepared to handle secure coding issues at work" (Q3). Although the rank is low, the average agreement is positive. The surprising result obtained for Q3 is likely related to the large number of neutral answers.
Further investigations are required to determine the root cause of this observation.

The collected results for RQ3 serve to inform practitioners who wish to design such games for an industrial context. It provides a ranked list of factors that participants consider to have a positive impact on CSC games. The three top factors that contribute to the success of a CSC game that should be considered by practitioners who wish refine the CSC game are the following: different kinds of challenges (Q17), based on real-life examples (Q20), and participants should work in teams rather and individually (Q16).

In terms of awareness, taking into consideration negative answers (-), neutral answers (N), and positive answers (+), the perception (PE), behavior (BE), and protection (PR) show the following results: PE$^-=8.04\%$, PE$^N=7.14\%$, PE$^+=84.82\%$, BE$^-=7.89\%$, BE$^N=20.79\%$, BE$^+=71.33\%$, PR$^-=7.78\%$, PR$^N=14.37\%$, PR$^+=77.84\%$.
These results show similar values for the negative answers (around 8\%), which might be related to the players' background. The highest result is related to perception, which also has the least amount of neutral answers.
While we observe strong agreement on the behavior and protection constructs (more than 70\%), there are still many neutral answers.
We believe that the large amount of neutral answers is also related to player background and the fact that the challenge type is not purely defensive, i.e., it is defensive/offensive, as discussed in section \ref{sec:csc_industry}.
The reasoning for this is based on the better results obtained in the study of the Sifu platform (see~\cite{Gasiba2020_CyberICPS}).

\subsection{Sifu/Online Cycle ––– CSC 10 to 13}
In this design cycle, the CSC challenges were further developed as the Sifu platform \cite{Gasiba2020_CyberICPS}.
The participants were asked to evaluate the platform through 5-point Likert scale questions.
Survey questions were based on the Awareness \cite{2014_Benenson_Defining_Security_Awareness}, and Happiness \cite{2018_Graziotin_Happy_Developers} dimensions.
The following is a summary of the results, in terms of the three awareness dimensions: perception (PE), behavior (BE), and protection (PR); and in terms of happiness (HP).
PE$^-=2.22\%$, PE$^N=8.89\%$, PE$^+=88.89\%$, BE$^-=0.0\%$, BE$^N=8.06\%$, BE$^+=91.94\%$, PR$^-=6.67\%$, PR$^N=11.11\%$, PR$^+=82.22\%$, HP$^-=8.22\%$, HP$^N=10.27\%$, HP$^+=81.51\%$.
The negative results (-) correspond to strongly disagree and disagree, neutral (N) to neutral answers, and positive results (+) correspond to agree and strongly agree.
A more in-depth analysis of these results, along with the Sifu platform's design, and the survey questions, can be found in \cite{Gasiba2020_CyberICPS_Journal}.
The collected answers again indicate an agreement with the awareness theory, in the following sequence: behavior, perception, and finally, protection.
Also, the participants report having fun and being happy while playing challenges in the Sifu platform.

The Hellinger distance is used to measure the distance between two probability mass functions (PMF).
The distance between the PMF of the three awareness constructs was computed to compare the results obtained in the second (refinement) and third cycle (Sifu/Online).
The obtained results are as follows (from higher distance value to smaller distance value): behavior ($d=0.25$), perception ($d=0.10$), and protection ($d=0.04$).
These results show that using the Sifu platform results in the most significant improvement in agreement on the behavior construct.
Although both cycles indicate positive results, the participants have a more substantial agreement that solving the Sifu platform's challenges helps in actual behavior (i.e., using defensive challenges), than using defensive/offensive challenges.
In terms of protection, the distance between the PMF is low (0.04), indicating that the agreement level is similar for the protection construct for both the defensive/offensive and the defensive challenges.
These results were as expected since the improvements to the challenges and the corresponding design cycles performed in the Sifu platform increase the adequacy to improve software developer awareness in terms of behavior.

\subsection{Discussions}
In this work, we have presented and evaluated an awareness training program for software developers in the industry, which was designed through three design cycles \cite{2004_hevner_design_science}.
The types of CSC challenges for each design cycle were as follows: offensive, defensive/offensive, and defensive.
The initial design cycle was mostly used for requirements elicitation to further develop and refine the CyberSecurity Challenges for software developers in the industry.
In the second design cycle, defensive/offensive challenges were introduced. These challenges adapt existing open-source projects to adopt a defensive perspective.
Finally, in the third design cycle, defensive challenges are introduced using the Sifu platform.
Our experience has shown that software developers highly appreciate playing CSC games based on direct feedback from participants.
It was also observed that playing CSC games can be done as either a standalone event or after a secure coding training.
Furthermore, the participants have claimed that the challenges have helped solidify, understand, and practice secure coding in real scenarios, the concepts discussed during training.
While the challenges, as described for the second and third design cycle, seem to address software developers and management's needs adequately, the third design cycle was shown to result in a higher agreement in terms of behavior.

Participants report on the happiness and fun in participating in these events.
However, a long term study on the impact of CSC events on software quality is not possible.
The reason for this is related to the large number of factors that hinder this study, which include, among others: job rotation, changing and evolving IT security technologies, discovery of new attack vectors, and evolving programming languages and programming language standards.
Therefore, we need to suffice with the fact that these events are both welcome by software developers and, with the fact that CSC has had continuous management approval throughout the years, and also the fact that it has been introduced in the standard teaching curriculum in the company where it was developed.

While previous work such as McIlwraith \cite{McIlwraith2006} provides a generic approach for awareness training, we show a method that explicitly addresses software developers in the industry and is based on a serious game inspired in the Capture-the-Flag format.
Nevertheless, some of the traits introduced by McIlwraith are also common with our artifact, e.g., the usage of web-based media and web-based text.
While the CSC artifact was designed for Web and C/C++ challenges, we think our approach can be generalized to other programming languages.
Other possible usages of our artifact include a refresher on previously acquired knowledge, a self-evaluation tool for individuals, and a recruiting tool used by human resources.
However, further work might be required either for non-industrial environments or participants with different backgrounds, e.g., management or human-resources.

\subsection{Threats to Validity}
There are threats to the validity of or findings - threats as they are typical or inherent to design research.
An example of these threats is the inclusion of mixed workshops in the study, which can introduce a bias in our analysis. While the authors cannot control the types of workshops, since they were dictated internal demand, we think that the obtained conclusions also apply to these workshops.
First, we evaluate the impact on awareness of IT-security topics.
The path from awareness training to secure products and services is long, and other research would be needed to evaluate whether such a game has impacted the quality of code.
Due to the large number of factors that affect code quality, this is, in practice, however not possible.
Nevertheless, awareness is a well-established endpoint in IT security research.
As in any design research, we cannot argue that our solution is the best, and we need to suffice with the argument that our artifact and outcome of research is successful, both in terms of developers happiness and management approval.
There are several external variables that we cannot control in an industrial setting that can limit our evaluations' validity.
Although we have explicitly mentioned to the participants that the survey questions refer to the CSC event, we cannot exclude questions' misinterpretation due to the participants' different cultural and language backgrounds.

Also, we cannot exclude a bias for socially desired answers and positive bias with the game setting.
However, for the validity of our findings, we refer to the fact that all game participants were industrial software engineers, and participation in the survey was not mandatory.
Our results demonstrate that these are a viable method for awareness training on secure coding in the industry in terms of the CSC game's usefulness.
We base this observation on the fact that it is approved by management, has high internal demand, and is liked and enjoyed by most participants.
%\vspace{-.4cm}
\section{Conclusions and Further Work}
\label{sec:conclusions}

In this work, we provide an overview of the design and implementation of CyberSecurity Challenges - a serious game to raise awareness on secure coding for software developers in the industry.
The CyberSecurity Challenges have been developed following a design science research design structured in three design cycles: Initial Design, Refinement, and Sifu/Online. The design cycles extended from 2017 until 2020 and consisted of thirteen events where more than 200 software developers participated.
Our contribution addresses practitioners who wish to develop or refine a software developer awareness training for the industry and the research community by understanding the usage of serious games targeting software developers in the industry.

This paper consists of two main parts: 1) an overview of the design of the CyberSecurity Challenges and 2) an evaluation of the CyberSecurity Challenge game and events, including the usefulness of CyberSecurity Challenges.
In the first part, we presented a consolidated view of CyberSecurity Challenges.
This consolidated view is the result of all the lessons learned throughout the three design cycles.
We provide an analysis and report of the main results that practitioners can use to design a similar awareness training program.
We also discuss the differences and similarities to other existing awareness training programs.
In the second part, we analyze results from semi-structured interviews from the first design cycle and a survey collected during the second design cycle.
Overall, software developers enjoy playing CyberSecurity challenges, either as a standalone event or together with a training workshop on secure programming.
Furthermore, we present results on the impact that the game has on the participants and discuss essential factors for successful awareness training.
Our positive results, continuous management endorsement, and the fact that these games have been introduced as a standard part of the company's teaching curricula validate our design approach.
Additionally, our results show that CyberSecurity challenges are a viable approach for awareness training on secure coding.

As further steps, the authors would like to design a systematic approach to identify topics for challenges and assessing these challenges for relevance. Towards this, more empirical analyses are required. Thus, parallel and next steps include an empirical study on the awareness of various secure coding topics to tailor the challenges to different software developer groups' needs. Also, as the COVID-19 crises limits travel and physical presence, we will continue to enhance the online version of the game. We also plan to enricht the scope of defensive challenges.

%\vspace{-.4cm}
\section*{Acknowledgements}
%xxx xxxxxxx xxxxx xxxx xx xxxxx xxx xxxxxxxxxxxx xx xxx xxxxxxxxxxxxx xxxxxxxxxx xxx xxxxx xxxx xxx xxxxx xxxxxxxx xxxxxxx xxx xxxxxxxxx
% xxxxxxxx xxxxxxx xxx xxxxxx xxxxxxxxxx
%xxxxx xxx xxxxxxx xxxxx xxxx xxxx xx xxxxx xxxx xxx xxxx xxx xxxxx xxxxxxxx xxxxxxxxxxx xxx xxxxxxxxxxxx xxxxxxxx xxx xxxxxxxxxxxx

The authors would like to thank the participants of the CyberSecurity Challenges for their time and their valuable answers and comments.
%% Kristian Beckers and Thomas Diefenbach
Also, the authors would also like to thank Kristian Beckers and Thomas Diefenbach for their helpful, insightful, and constructive comments and discussions.

%%%%%%%%%%%%%%%%%%%%%%%%%%%%%%%%%%%%%%%%%%%%%%%%%%%%%%%%%%%%%%%%%%%%
%%%% NOTE: this shall be enabled in the final version - since this is a double-blind review, I think it must be disabled in the version for review
%%%%%%%%%%%%%%%%%%%%%%%%%%%%%%%%%%%%%%%%%%%%%%%%%%%%%%%%%%%%%%%%%%%%
This work is financed by portuguese national funds through FCT - Fundação para a Ciência e Tecnologia, I.P., under the project FCT UIDB/04466/2020. Furthermore, the third author thanks the Instituto Universitário de Lisboa and ISTAR-IUL for their support.
%%%%%%%%%%%%%%%%%%%%%%%%%%%%%%%%%%%%%%%%%%%%%%%%%%%%%%%%%%%%%%%%%%%%

%xxxx xxxx xx xxxxxxxx xx xxxxxxxxxx xxxxxxxx xxxxx xxxxxxx xxx x xxxxxxxx xxxx x xxxxxxx x xxxxxxxxxxx xxxxx xxxxx xxx xxxxxxx xxx xxxx xxxxxxxxxxx. xxxxxxxxxxxx xxx xxxxx xxxxxx xxxxxx xxx xxxxxxxxx xxxxx xxxxxxxx xx xxxxxx xxx xxxxxxxxx xxx xxxxx xxxxxxxx

%\scriptsize
\footnotesize
\small
%\setlength{\bibsep}{0ex}
%\vspace{-.4cm}
\bibliographystyle{splncs04}
\bibliography{bibliography}

\end{document}